\documentclass[aps, pre, preprint]{revtex4-1}
\usepackage{graphicx, amssymb, amsmath, enumerate, placeins, bm, color, soul}
\usepackage[english]{babel}
\usepackage{url}
\usepackage[colorlinks, allcolors = blue, bookmarksopen,bookmarksnumbered]{hyperref}

\newcommand{\ngo}{\textit{N. gonorrhoeae }}
\newcommand{\ngows}{\textit{N. gonorrhoeae}}
\newcommand{\vb}{\mathbf{v}}
\newcommand{\vd}{\overline{\mathbf{v}}_d}
\newcommand{\rb}{\mathbf{r}}
\newcommand{\A}{\mathcal{A}}
\newcommand{\B}{\mathcal{B}}

\begin{document}
    \title{Rectification of twitching bacteria through narrow channels: A numerical simulations study}
    \date{\today}
    \author{Konark Bisht}
    \email{Konark.Bisht@physics.iitd.ac.in}
    \author{Rahul Marathe}
    \email{maratherahul@physics.iitd.ac.in} 
    \affiliation{Department of Physics, Indian Institute of Technology Delhi, New Delhi 110016, India.}
    
    \begin{abstract}
        Bacteria living on surfaces use different types of motility mechanisms to move on the surface in search of food or to form micro-colonies. Twitching is one such form of motility employed by bacteria such as \ngows, in which the polymeric extensions known as type IV pili mediate its movement. Pili extending from cell body adheres to the surface and pulls the bacteria by retraction. The bacterial movement is decided by the two-dimensional \textit{tug-of-war} between the pili attached to the surface. Natural surfaces in which these \textit{micro-crawlers} dwell are generally spatially inhomogeneous and have varying surface properties. Their motility is known to be affected by the topography of the surfaces. Therefore, it is possible to control bacterial movement by designing structured surfaces which can be potentially utilized for controlling biofilm architecture. In this paper, we numerically investigate the twitching motility in a two-dimensional corrugated channel. The bacterial movement is simulated by two different models: (a) a detailed \textit{tug-of-war} model which extensively describe the twitching motility of bacteria assisted by pili and (b) a coarse-grained \textit{run-and-tumble} model which depicts the motion of wide-ranging self-propelled particles. The simulation of bacterial motion through asymmetric corrugated channels using the above models show rectification. The bacterial transport depends on the geometric parameters of the channel and inherent system parameters such as persistence length and self-propelled velocity. In particular, the variation of the particle current with the geometric parameters of the micro-channels show that one can optimize the particle current for specific values of these parameters.
    \end{abstract}
    
    \keywords{Twitching motility, Ratchet effects; Rectification; Active particles}
    
    \maketitle
    \section{Introduction}\label{sec:1}
    
    Many bacterial species dwell on surfaces. Each species employ a particular or a combination of different motility mechanisms such as swimming, darting, gliding and twitching to survey the surfaces for food and colonies \cite{RM2003}. Surface-dwelling bacteria such as \textit{Neisseria gonorrhoeae} cannot actively swim in a liquid medium; instead, they possess a form of surface motility known as ``\textit{twitching motility}'' \cite{JH1983, JM2002}. It is mediated by the polymeric extensions emerging from the cell body known as \textit{pili} and is characterized by intermittent and jerky movement \cite{JS2001, CG2012}. Pili have various functions, but a class of these pili known as type-IV pili (T4P) assist in motility over surface \cite{JH1983, JM2002, BM2015}. T4P undergoes cycles of polymerization and depolymerization. The polymerization results in the generation of new pilus and subsequent elongation, out of which some get attached to the surface, whereas, depolymerization leads to retraction and eventual disappearance of a pilus. The retracting pilus, attached to the surface, exerts a force on the cell body and pulls the bacterium along its direction. A single pilus can generate a force exceeding $100$ pN \cite{AM2000, BM2002, NB2008}. The vectorial balance of forces exerted by various T4P in different directions as in a \textit{tug-of-war} eventually decides the direction of motion of the bacterium \cite{MM2008, CH2010, RM2014, VZ2014}. 
    
    Unlike many other bacterial species, the \ngo cannot sense the chemical gradient as the chemotactic gene is absent in them \cite{JT2015}. Due to the absence of any biases, these ``\textit{micro-crawlers}" essentially executes a persistent random walk on flat surfaces. In its natural habitat, however, the bacteria encounter surfaces having spatial inhomogeneities and different surface properties. The topography of the surface influence the motility of the bacteria as recent experimental studies indicate that \ngo can sense the topography of the surfaces and the microscopic structures can guide their movement \cite{CH2009, CM2012}. It has led to the designing of structured surfaces where the physio-chemical properties can control the bacterial movement and potential biofilm architecture \cite {CH2009, BM2013}. In general, controlling microbial locomotion have potential applications in diverse areas such as diagnostics \cite{PD2012},  therapeutic protein synthesis \cite{BR2010}, photosynthetic biofuel production \cite{JG2008, SS2009, EE2011} and microfluidic devices \cite{MK2007, MK2008}.
 
One of the way to control the bacterial movement and have directed motion is by exploiting the so-called \textit{active ratchets} \cite{CR2002, LA2011, YY2011, CR2013, JWu2015, YF2015, BH2015,  BA2017, CR2017}. They are realized by the self-propelled particles (SPPs) which can be biological (e.g. bacteria) or non-biological (e.g. Janus particles) in nature \cite{SH2008, GM2009, PG2013, JK2018}.

SPPs are mobile agents which convert energy from an environment into persistent motion and are referred to as the active matter. In these ratchets, the SPPs moving through an asymmetric medium show rectification in their motion and there is a net transport of particles in a particular direction \cite{PG2013, XA2014, JW2014, JW2015, CR2017, SP2019}. This asymmetrical or directed response can be accounted to the spatial and temporal asymmetry in the system \cite{PR2002, LA2011, SD2014}. The temporal asymmetry is inherent in the motion of SPPs as the presence of self-propelled velocity with orientational fluctuations drives the system out-of-equilibrium. On the other hand, the spatial asymmetry is imposed by the inhomogeneities in the medium. The boundaries of the channel limit the phase space volume of the particle resulting in the emergence of an effective entropic potential which reflects the variations in the cross-section of the channel. The modulation of the cross-section with a periodic boundary function having broken reflection symmetry induces a symmetry breaking which biases the particle motion \cite{PR2002, PM2013}.
    
Rectification has been observed in the motion of biological micro-swimmers such as \textit{Escherichia coli}, in a chamber with an array of funnel barriers \cite{PG2007, PG2008, GL2010, YC2015}, or through ratcheting micro-channels \cite{SH2008, JR2016}. The directional rotation of gears with asymmetric teeth in a suspension of bacteria \cite{AS2010, RL2010} or in the presence of self-propelled robots \cite{HL2013} is another such example of rectification process in active matter systems due to spatial asymmetry. It holds significance in the context of using bacterial suspension to power mechanical micro-machines \cite{YH2006, AG2008, LA2009, SK2016,  AS2018, AS2019}. However, similar studies on the transport of surface motile twitching bacteria in the asymmetric medium are lacking. Unlike the micro-swimmers, the hydrodynamic interactions are absent for these micro-crawlers, and only steric surface interactions could play a role in the rectification process.
    
    In this paper, we numerically investigate the twitching motility of bacteria in narrow two-dimensional (2D) channels having periodic boundaries with broken reflection symmetry. The stochastic \textit{tug-of-war} model (TWM) is implemented to mimic the twitching motility of \ngo \cite{RM2014}. To characterize the motion,  mean squared displacement (MSD) is computed, which identify distinct diffusion regimes for different time scales. The bacterium confined in the corrugated channel experience rectification in motion as evident by the non-zero value of mean displacement along the axis of the channel. The particle current which gives a measure of net transport of bacteria is dependent on the relative value of the persistence length to the size of the compartment of the channel. It is observed that the particle current can be optimized by careful selection of geometric parameters of the channel. We also compare our results with a coarse-grained \textit{run-and-tumble} model (RTM), which is a generic model used to describe the motion of SPPs \cite{JH2007, HB2014}.
    
    The rest of the paper is organized as follows. In Sec.~\ref{sec:2}, the two models of twitching motility are introduced. In Sec.~\ref{sec:3}, we first discuss the geometry of the corrugated channel used in the simulation. Next, in the section, relevant quantities such as MSD, mean displacement along the $x$ axis and spatial probability density, velocity profiles are evaluated to investigate the rectified motion of a bacterium. In the latter part, the dependence of particle current on geometric parameters of the channel for the two models is explored. Finally, in Sec.~\ref{sec:4}, we conclude by providing a summary and discussion of our results.
    
    \section{Models for twitching motility}\label{sec:2}
    \subsection{Twitching motility using stochastic \textit{tug-of-war} model}\label{sec:2a}
    
    We simulate the twitching motility of \ngo bacteria using the 2D stochastic \textit{tug-of-war} model (TWM) described in Ref.~\cite{RM2014}. The cell body of the bacterium is modeled as a point particle with straight rodlike pili emerging radially from the cell body in random directions. A pilus is considered to be in one of the three states: elongating, retracting or attached to the surface states. The pilus stochastically switches between these states with the rates that are estimated from the experiments. The retraction velocity and the unbinding rate of a pilus depend on the force experienced by the pilus. When the number of pili in opposite directions are different, the pilus on the side with lesser number of pili experience greater force than on the opposite side. Hence, it is more likely to unbind from the surface due to the force-dependent nature of the unbinding rate. The unbinding of pilus in the weaker side further increases the imbalance causing a sharp escalation of unbinding of pili from the weaker side resulting in a rapid motion in the direction of pili who win the \textit{tug-of-war}. The further rebinding or unbinding of pili, dissolution of pili due to full retraction or creation of new pili, alter the force balance resulting in the change in the direction of motion. 
    
  The one-dimensional (1D) tug-of-war models and two-state models have been applied to study the dynamics of molecular motors \cite{MM2008, TG2011, PMalgaretti2017}. However, there are quantitative differences in the dynamics in a 1D model than the 2D TWM used in this work. In the 1D models of molecular motors, the cargo gets pulled in two opposite directions only. This would correspond to pili extending in two opposite directions in case of twitching bacteria. However, in case of \ngo  where pili extend in all directions $[0, 2\pi)$. Therefore, it requires a 2D tug-of-war mechanism to simulate the twitching motility in \ngows. Our model also incorporates directional memory by allowing the pilus bundling and the re-elongation of fully retracted pili with some finite probability \cite{RM2014}. These two additional properties in the 2D model are essential to mimic the experimentally observed increase in the persistence time of \ngo with the increasing pilus number \cite{CH2010, RM2014}. The persistence time $t_p$ is the average time for which the overall direction of motion remains same. The persistence time $t_p$ is proportional to the persistence length $\ell_p$ as $t_p=\ell_p/v$, where $v$ is average speed of the bacterium. On the other hand, the average number of pili increases with the rate of pilus creation $R_{cr}$.  Therefore, it can be stated that the persistence length (or time) increases with the rise in the pilus creation rate as also seen in experiments \cite{CH2010}.  In a wild type \ngo bacterium, a persistent length of $1-2$ $\mu$m corresponds to average seven pili \cite{CH2010, RM2014}. The trajectories of different persistence length are obtained by varying the $R_{cr}$ in our simulation using TWM. 
    
    Since our interest is in the study of the twitching motility in the confined narrow channels, the nature of motility near the boundaries plays a significant role in the transport properties. The boundary condition in TWM is implemented in the following manner. When the bacterium is near the boundary, pili cannot attach to surface beyond the boundary. So the pili can attach only in a direction which is towards the interior of the channel. It restricts the bacterium to take a step beyond the boundaries and is thus confined in the channel.
    
    \subsection{Twitching motility as \textit{run-and-tumble} motion}\label{sec:2b}
    
    In this section, we discuss the motivation towards the coarse-grained model of the twitching motility. The motility of \ngo have been recorded and analyzed in various experimental studies \cite{CH2010, CM2012, BM2013}. In these experiments, the bacteria crawling on glass plates were observed under a microscope, and movies tracking their positions were recorded for a duration of a few minutes. A distribution of step lengths $\ell$ was generated by analyzing the recorded tracks \cite{CH2010, KB2017}. The step length distribution $P(\ell)$ was found to follow an exponential function given by
    \begin{equation}
        P(\ell) = \frac{1}{\ell_p}\exp(-\ell/\ell_p).
        \label{eq:1}
    \end{equation}
    The persistence length $\ell_p$ is the average distance a bacterium travels before taking a turn. A typical value of the average speed $v$ of \ngo was reported to be 1.5--$2~\mu$ms$^{-1}$ \cite{CH2009}. A coarse-grained model for bacterial motility can be constructed using experimentally observed features of \ngo walks \cite{KB2017}. The bacterium is modeled as a point particle executing a 2D random walk with step lengths drawn from the exponential distribution of Eq.~(\ref{eq:1}). The particle selects a new direction at each turn from a uniform distribution between $[0, 2\pi)$. A fixed value of average speed $v=1.5~\mu$ms$^{-1}$ is taken with the time duration to complete a step of $\ell$ length is given by $\ell/v$. The motility is characterized by straight trajectories with sudden random changes in the direction. The resulting motion is referred to as \textit{run-and-tumble} and is quite ubiquitous in SPPs. In our model, the time duration of the tumbling event is zero, and change in the direction happens instantaneously at each turn.
    
    Using the above model of motility, we study the transport of bacteria in a corrugated channel. The boundary condition is taken in which the bacterium is reflected towards the interior in a random direction. So, when the bacterium hits the boundary wall, a new proposed direction is selected randomly from a uniform distribution [0, $2\pi$). If it is not towards the interior, another direction is chosen until the proposed direction is towards the interior of the channel. The bacterium instantaneously turns to the new direction and is hence prevented from taking steps beyond the boundary walls.
    
    \section{Simulation methods and results}\label{sec:3}
    
    \begin{figure}[!htbp]
        \centering
        \includegraphics[scale = 0.5]{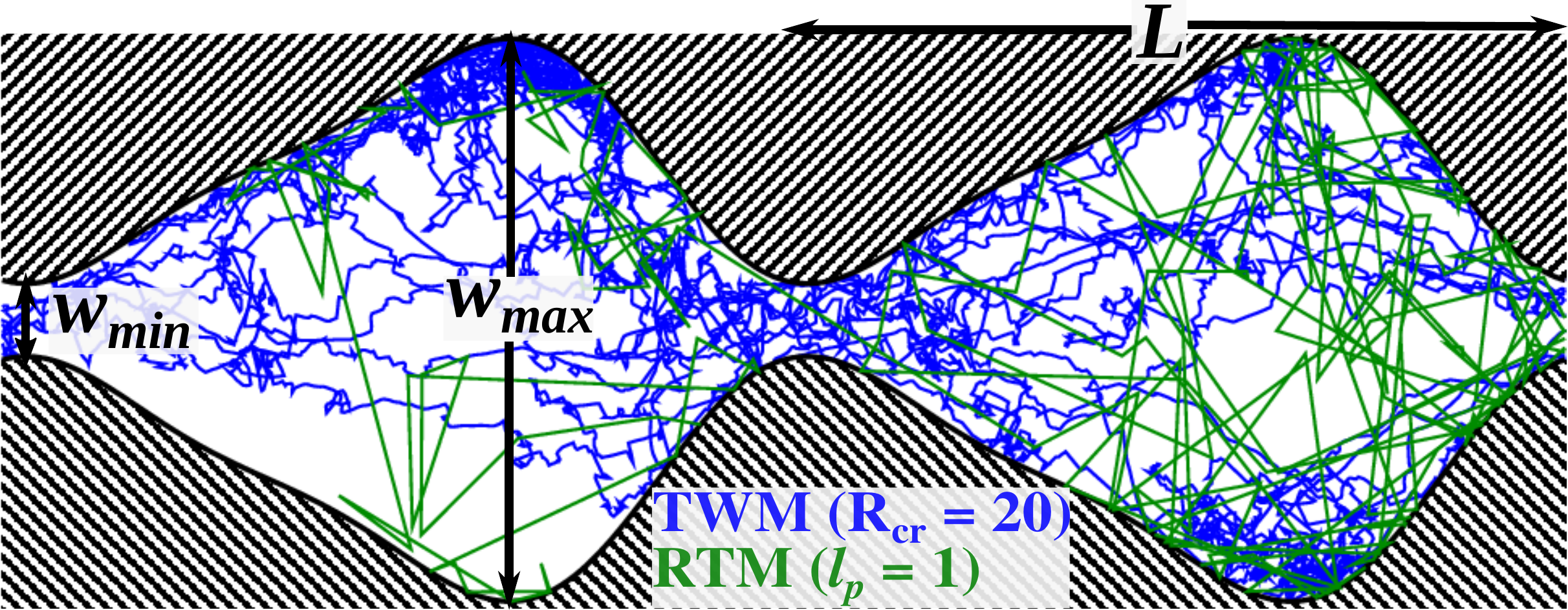}
        \caption{Tracks obtained from simulation of bacterial motility in corrugated channel using (i) TWM (for $R_{cr} = 20$ s$^{-1}$) and (ii) RTM (for $\ell_p =1$ $\mu$m).}
        \label{fig:1}
    \end{figure}
    We simulate the twitching motility using the two models described in section \ref{sec:2} The bacterial motion is confined in a 2D corrugated channel which extends infinitely along $x$ axis and is bounded by a periodic function along $y$ axis (Fig.~\ref{fig:1}). Boundaries of the asymmetric channel are modeled by a boundary function having broken reflection symmetry which is widely used in literature \cite{RB1994, PJ1996, RB1997, PR2002, PMJCP2013, PH2009}. The boundary function $w(x)$ is given by,
    \begin{equation}
        w(x) = \A-\B\left[\sin\left(2\pi \frac{x}{L}\right) + \frac{\Delta}{4}\sin \left(4\pi \frac{x}{L}\right)\right].
        \label{eq:2}
    \end{equation}
    Here, $L$ is the spatial periodicity along $x$ axis of the channel. The extent of asymmetry in the shape of the channel is determined by the asymmetric parameter $\Delta$. For a symmetric periodic channel, $\Delta = 0$. The channel is biased towards the positive $x$ direction for $\Delta < 0$ and the negative $x$ direction for $\Delta > 0$. In this study, the simulations are primarily done for a channel biased towards the negative $x$ direction. The parameter $\A$ determines the half-width of the bottleneck whereas $\B$ controls the slope of the channel boundaries \cite{AB2009}. The parameters $\A$ and $\B$ are related to the minimum width $w_{min}$ and the maximum width $w_{max}$ of the channel as
    \begin{equation}
        w_{min} = 2(\A-\B\delta)\mbox{ and }w_{max} = 2(\A+\B\delta).
        \label{eq:wmin}
    \end{equation}
    where $\delta$ is obtained by finding the point of extremum of the boundary function $w(x)$ in Eq.~(\ref{eq:2}) and is given by,
    \begin{equation}
        \delta(\Delta) = \left[\left({\frac{\Delta^2-1+\sqrt{1+2\Delta^2}}{2\Delta^2}}\right)^{1/2} +\frac{1}{4}\left(\frac{\Delta^4-2(\Delta^2-\sqrt{1+2\Delta^2}+1)}{\Delta^2}\right)^{1/2}\right].
        \label{eq:delta}
    \end{equation}
    From Eq.~(\ref{eq:delta}), we can note that the $\delta$ is an monotonic increasing function of $\Delta$ for $\Delta>0$. For a typical value of $\Delta = 1$, $\delta = 1.1$. Simulations are performed for the parameter values $\Delta = 1$, $\A = 1~\mu$m, $\B = 0.7~\mu$m and $L=5~\mu$m unless explicitly mentioned otherwise. For these values of the parameters, $w_{max} \simeq 3.5~\mu$m and $w_{min} \simeq 0.46~\mu$m. The ensemble averages of all physical quantities is denoted by $\langle..\rangle$ are derived by averaging over $10^4$ trajectories unless mentioned otherwise. In our simulations, the bacterium always starts from a fixed initial position $(x_0, y_0) = (0,0)$ at time $t=0$.
    
    \begin{figure}[!htbp]
        \centering
        \includegraphics[scale = 0.1]{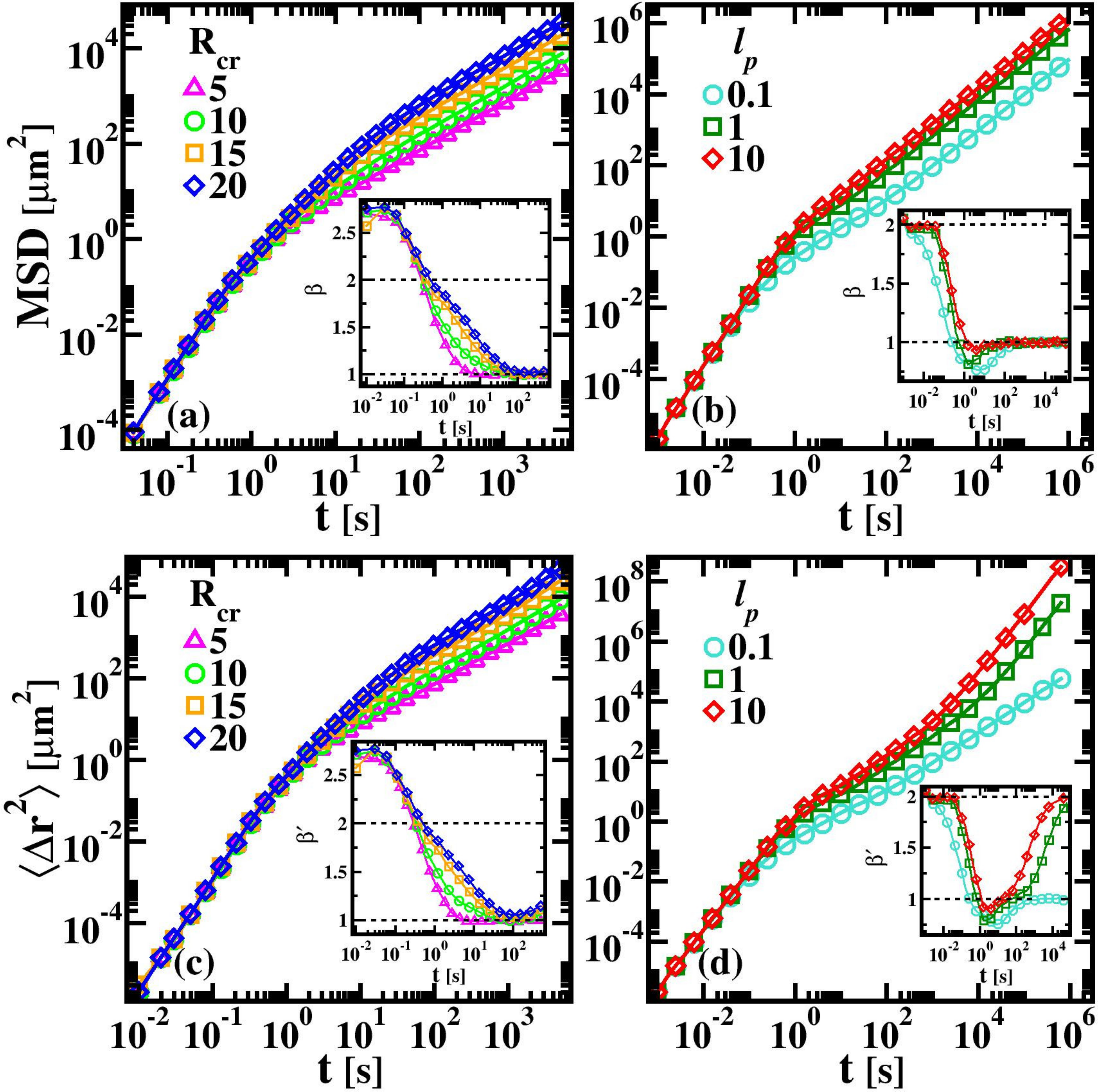}
        \caption{(a) MSD $\langle\Delta\rb^2\rangle - \langle\Delta\rb\rangle^2$ as a function of time $t$ for $R_{cr} =5$, 10, 15, 20 s$^{-1}$ obtained by simulation of the bacterial motility in the corrugated channel using TWM. (b) The plot of $\langle{\Delta\rb}^2\rangle$ vs. $t$ for $\ell_p = 0.1$, 1, $10~\mu$m for RTM. $\langle\Delta\rb^2\rangle$ as a function of  $t$ for  (c) TWM and (d) RTM. Insets: $\beta$ as a function of $t$ in (a) and (b), and  $\beta^\prime$ as a function of $t$ in (c) and (d).}
        \label{fig:r2}
    \end{figure}

 The prototypical tracks traced by a bacterium having $R_{cr} = 20$ s$^{-1}$ and $\ell_p =1$ $\mu$m, simulated by TWM and RTM, respectively are shown in Fig.~{\ref{fig:1}}. The tracks from TWM are detailed and capture the jerky movements of the bacterium. On the other hand, tracks from RTM lacks the detailed structure and gives a coarse-grained representation of the motility. In both the cases, the movement is directed at short times and random at longer time scales. It is demonstrated by the plot of the mean squared displacement (MSD) $\langle\Delta{\rb(t)}^2\rangle-\langle\Delta\rb(t)\rangle^2$ as a function of time $t$ in Figs.~\ref{fig:r2}(a) and \ref{fig:r2}(b), where $\Delta\rb(t) = \rb(t)-\rb(0)$. The dynamics is characterized by MSD exponent $\beta$ defined by MSD $\sim t^\beta$.  Fig.~\ref{fig:r2}(a) show the plot of MSD calculated from the trajectories of TWM for $R_{cr}=5$, 10, 15, 20 s$^{-1}$. In the inset, the variation of $\beta$ with $t$ is shown which is evaluated from the relation $\beta(t) = \log_{10}[\langle\Delta\rb^2(10t)\rangle/\langle\Delta\rb^2(t)\rangle]$. The $\beta \simeq 2$ for short time scales indicates ballistic motion. The exponent gradually reduces to $\beta \simeq 1$ indicating diffusion for longer time scales. In TWM, the motion achieves the steady state after some period of time. In this initial transient state of the motion, $\beta>2$ for very small $t$. Fig.~\ref{fig:r2}(b) depict the plot of MSD computed from RTM trajectories for $\ell_p= 0.1$, 1, $10~\mu$m. The motion changes from super-diffusive at very short time scales ($t\lesssim 0.1$ s) to sub-diffusive at intermediate time scales ($1\lesssim t\lesssim 100$ s). The motion is diffusive for longer time scales ($t\gtrsim 100$ s). We do not see much difference in the dynamics of the twitching bacteria in the channel geometry than in an unconfined surface. However, when we plot of $\langle\Delta\rb(t)^2\rangle$ with $t$, we see anomalous motion for long times. The variation of $\langle\Delta\rb(t)^2\rangle$ has been employed to search for signatures of rectification in the dynamics of SPPs in Refs.~\cite{AC2015, SP2019, AW2019}. In Figs.~\ref{fig:r2}(c) and \ref{fig:r2}(d), we plot the $\langle\Delta\rb^2\rangle$ vs $t$ for TWM and RTM, respectively. The insets show the variation of exponent $\beta^\prime$ with  $t$ which is defined by the relation $\langle\Delta\rb^2\rangle\sim t^{\beta^\prime}$. We observe that the exponent $\beta^\prime >1$ at larger time scale in this case for bacterium having $\ell_p$ comparable to the channel dimensions which is distinct from the behavior of  MSD exponent $\beta\simeq1$ for large $t$ for all values of $\ell_p$. This divergence is due to the non-zero finite value of $\langle\rb\rangle$ at long times due to rectification which is accounted in the MSD calculations but not in the case of $\langle\Delta\rb^2\rangle$.

    \begin{figure}[!htbp]
        \centering
        \includegraphics[scale = 0.095]{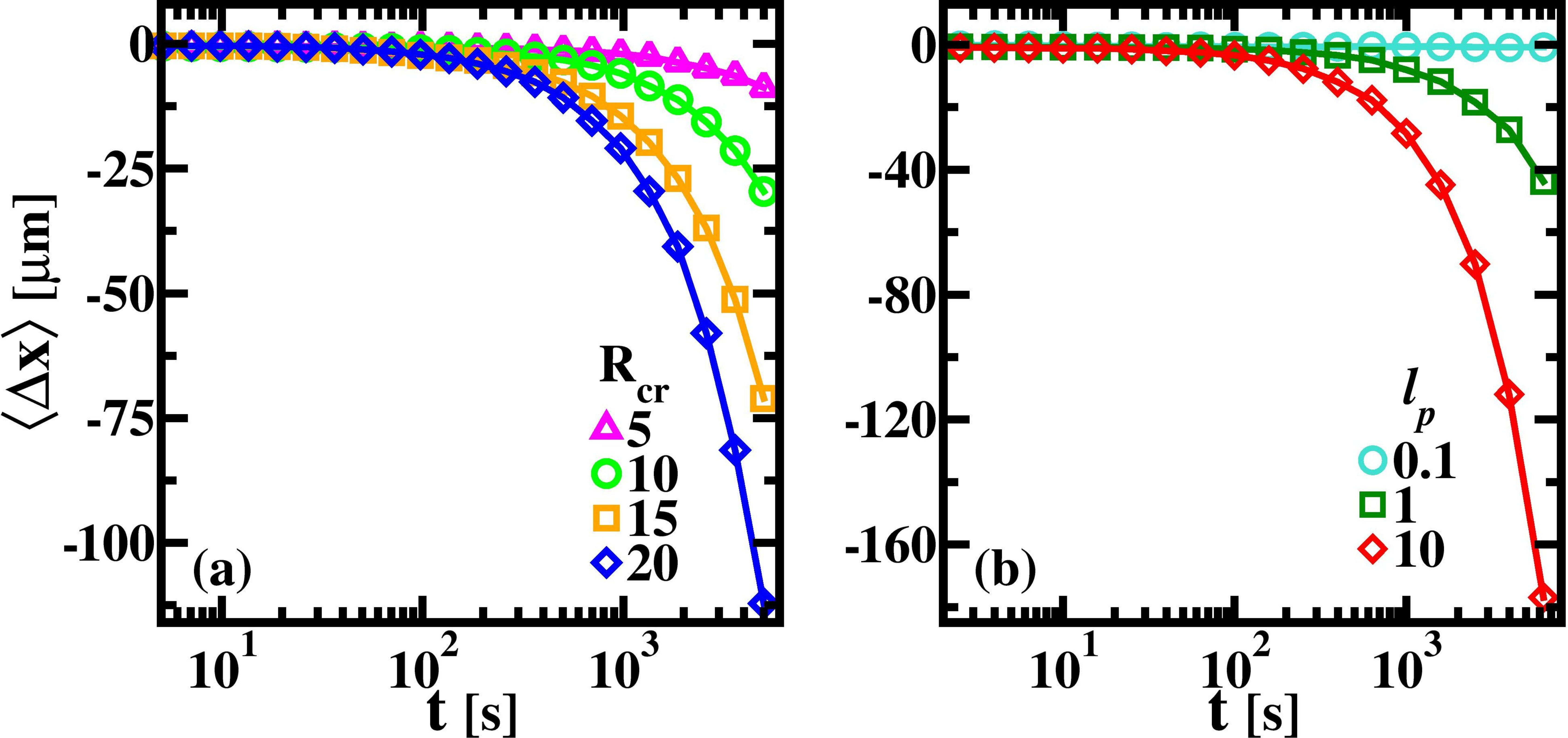}
        \caption{Mean displacement along $x$ axis $\langle\Delta x\rangle$ vs $t$ using (a) TWM for $R_{cr} =5$, 10, 15, 20 s$^{-1}$, and (b) RTM for $\ell_p = 0.1$, 1, $10~\mu$m.}
        \label{fig:mx}
    \end{figure}
    In order to ascertain the rectification in bacterial motion, we compute the mean displacement along $x$ axis $\langle\Delta x(t)\rangle$ where $\Delta x(t) = x(t)-x(0)$. The plot of $\langle\Delta x(t)\rangle$ vs $t$ obtained for two models TWM [in Fig.~\ref{fig:mx}(a)]  and RTM [in Fig.~\ref{fig:mx}(b)] show finite displacement $\left(\vert\langle\Delta x(t)\rangle\vert\neq 0\right)$ for large time scales and it increase with the persistence length in both models. The plots are steep for larger values of persistence length which indicates that the particle current may have a dependence on the persistence length. It is explored in detail in the later part of this section.

   \begin{figure}
        \centering
        \includegraphics[scale=0.32]{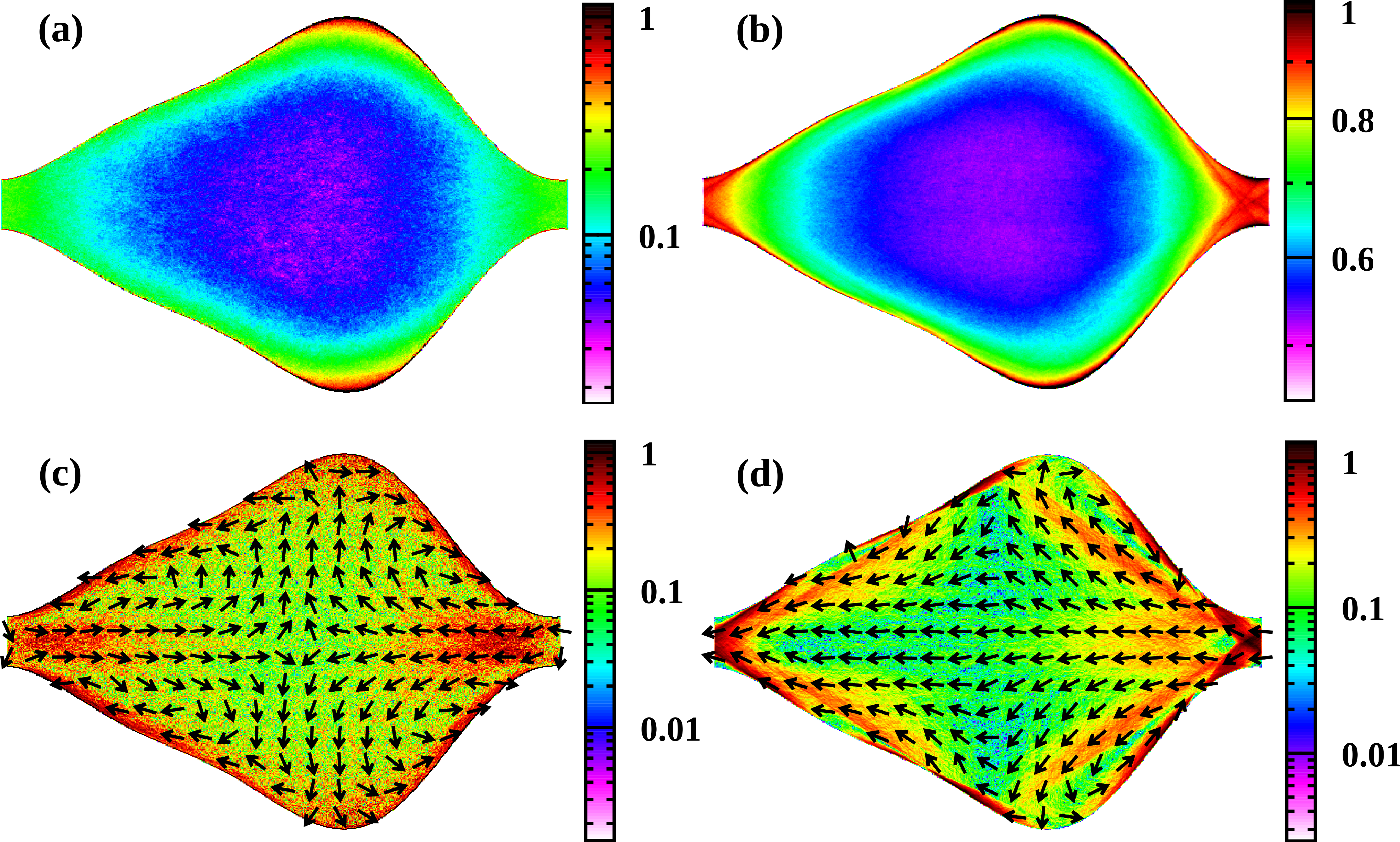}
        \caption{Top row shows the color map of scaled spatial probability density $\rho(\rb)$ for (a) $R_{cr} = 10$~s$^{-1}$ in TWM and (b) $\ell_p =10~\mu$m in RTM. Bottom row depicts the corresponding profiles of average drift velocity $\overline{\vb}(\rb)$ for (c) TWM and (d) RTM where color maps shows the magnitude of drift velocity profile $\overline{\vb}(\rb)$ and the arrows depicts its direction.}
        \label{fig:rhoV}
    \end{figure}
    The directed transport in the asymmetric channels is further investigated by calculating the spatial probability density $\rho(\rb)$  and time-averaged drift velocity $\overline{\vb}(\rb)$ profiles from the bacterial trajectories obtained from simulations for the time duration $t=10^6$ s. Figs.~\ref{fig:rhoV}(a) and \ref{fig:rhoV}(b) depict the colour maps of $\rho(\rb)$ for (a) TWM (for $R_{cr} =  20$~s$^{-1}$) and (b) RTM (for $\ell_p=10~\mu$m). The $\rho(\rb)$ values are scaled by its largest values in the compartment. The $\rho(\rb)$ profile show that the bacterium tend to spend more time near the boundaries for larger $R_{cr}$. This was also observed experimentally when \ngo bacteria were allowed to interact with 3D obstacles \cite{CM2012}. It should also be noted that the twitching bacteria tend to accumulate at places where the channel cross-section is widest. The same observation was also reported for different kinds of SPPs in the theoretical study using Fick-Jacobs approximation in Ref.~\cite{PM2017}. Since bacterium having large persistence length ($R_{cr}= 20$ s$^{-1}$ or $\ell_p$ = 10) stays near the boundaries, it experiences the asymmetric shape of the channel more often which causes an overall directed motion along the channel. On other hand, for small persistence lengths ($\ell_p=0.1~\mu$m) bacteria encounters the boundary rarely and move randomly in the interior of the channel. 

The average drift velocity $\vd$ fields are plotted in Figs.~\ref{fig:rhoV}(c) and \ref{fig:rhoV}(d). The color map shows the scaled magnitude of time-averaged drift velocity profile $\vd(\rb)$ in the compartment. The magnitude of $\vd(\rb)$ is computed as $\vert\vd(\rb)\vert = \left(\vd(\rb)_x^2+\vd(\rb)_y^2\right)^{1/2}$ which is then scaled with its largest value in the compartment. The arrows in Figs.~\ref{fig:rhoV}(c) and \ref{fig:rhoV}(d) depicts the normalized average drift velocity profiles $\hat{\vd}(\rb) =\vd(\rb)/\vert\vd(\rb)\vert$.  In Fig.~\ref{fig:rhoV}(c), the velocity profiles for TWM  are shown which indicates the bacteria tend to diffuse away from constriction. On the other hand, in Fig.~\ref{fig:rhoV}(d), the bacteria undergoing run-and-tumble move towards the direction of biasing of the channel for $\ell_p>L$.
    
    An essential quantity that quantifies the particle transport across the channel is the particle current. Since the geometry of our system is a quasi-1D along the $x$ axis, the particle current $J_x$ can be considered to be due to motion along the $x$ axis alone and is defined by \cite{PR2002, GS2009, BA2009}, 
    \begin{equation}
        J_x = -\lim_{t\to\infty}\frac{\langle\Delta x(t)\rangle}{t}.
        \label{eq:Jx}
    \end{equation}
    In our work, we have defined our channel to be biased toward negative $x$ direction in most cases. Therefore, we have taken a negative sign in our definition of $J_x$ so that the $J_x\gtrsim 0$ for a channel biased toward $-x$ direction which helps in visualization. The current  $J_x <0$ for a channel biased toward $+x$ direction. 

    In all our simulations, $J_x$ is calculated at time $t = 10^3$ s. This time is sufficient time to reach a steady-state such, and the current is stabilized. Due to the effect of orientational fluctuations intrinsic to self-propulsion mechanism, the bacteria changes its direction randomly, which results in particle current to be only a small fraction of the self-propelled speed $v$. The magnitude of current may depend on a number of intrinsic factors, such as persistence length or self-propelled speed, and on the geometric parameters of the channel. 
    
    \begin{figure}[!htbp]
        \centering
        \includegraphics[scale = 0.08]{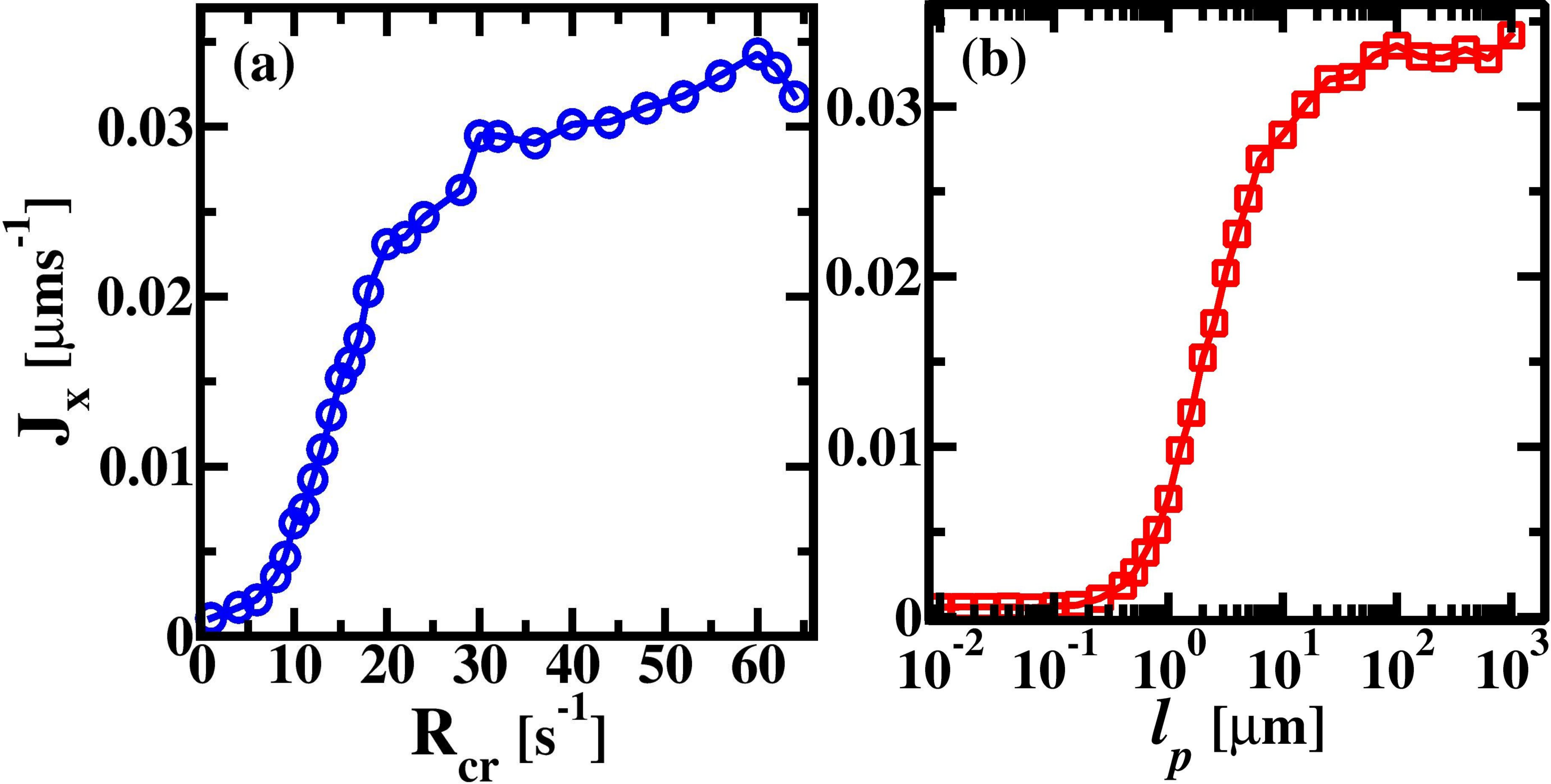}
        \caption{(a) The variation of the particle current $J_x$ computed for TWM with the pilus creation rate $R_{cr}$. (b) The plot of $J_x$ with the persistence length $\ell_p$ for RTM.}
        \label{fig:JxVsRcrlp}
    \end{figure}
    Next, we study the dependence of the particle current on the persistence length. In TWM, the persistence length is directly proportional to the pilus creation rate $R_{cr}$ whereas in RTM, $\ell_p$ determines the persistence of the walks (see Sec.~\ref{sec:2}). In Fig.~\ref{fig:JxVsRcrlp}(a), we show the plot of the variation of $J_x$ with $R_{cr}$ and found $J_x$ to increase with the rise in $R_{cr}$.  It is understandable since as $R_{cr}$ increase, the average number of pili increase, which in turn leads to an increase in persistence length. The bacterium encounters the boundary walls more for larger persistence length resulting in enhanced $J_x$. The current tends to approach a maximum value for a larger value of $R_{cr}$. Fig.~\ref{fig:JxVsRcrlp}(b) show the plot of  $J_x$ vs $\ell_p$ where we have taken a large variation of $\ell_p$. The $J_x\simeq 0$ for small values of $\ell_p$ ($\ell_p\lesssim 0.5~\mu$m) as bacterium rarely experience the asymmetric boundaries. The $J_x$ increases rapidly in the intermediate region where the $\ell_p$ is comparable to the dimension of the compartment {$0.5\lesssim \ell_p\lesssim 10~\mu$m}. The current then becomes independent of $\ell_p$ and saturates for larger values of $\ell_p$ ($\ell_p\gtrsim 100~\mu$m). For such a high value of persistence length ($\ell_p>>L$), the motion is increasingly restricted by the compartment walls, which makes the further increase in the persistence length ineffective. Therefore, as there is no effective increase in persistence, there is no further increase in the current. It indicates that the particle current has a strong dependence on the relative value of persistence length to that of the geometric parameters of the channel.
 We have also explored the dependence of particle current $J_x$ on the self-propulsion velocity $v$ in RTM. The particle current $J_x\sim v$ for large values of $\ell_p$ for sufficiently large $v$.
    
    \begin{figure}[!htbp]
        \centering
        \includegraphics[scale = 0.08]{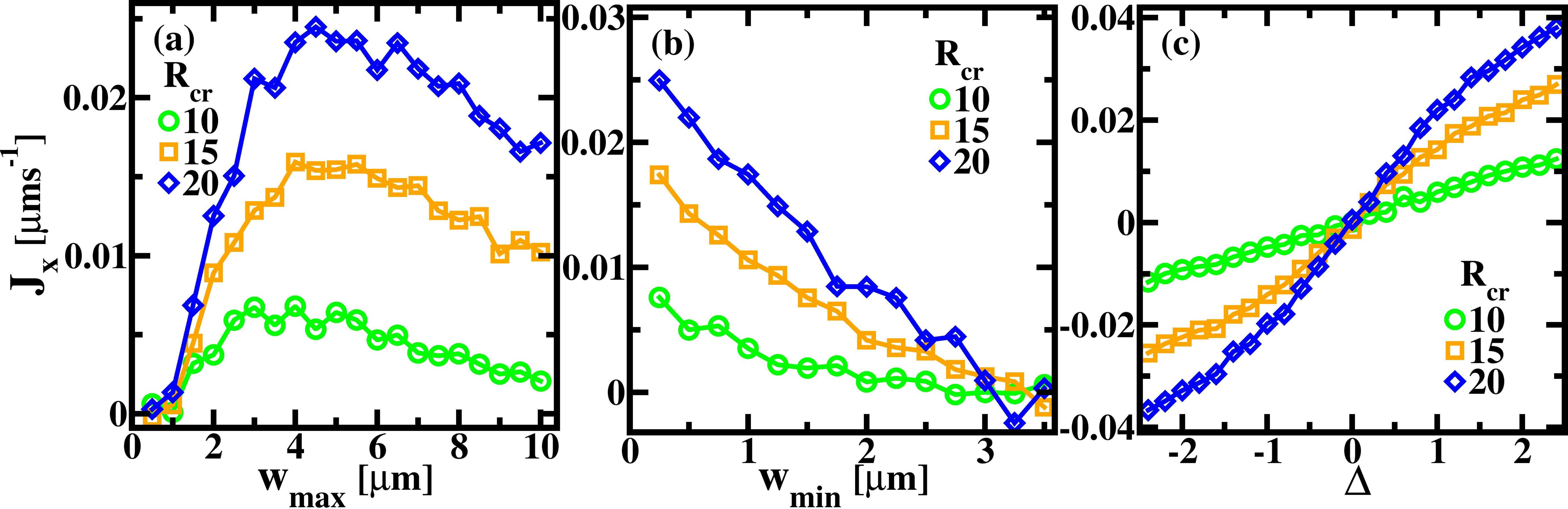}
        \caption{The dependence of the particle current $J_x$ with the channel parameters: (a) maximum width $w_{max}$, (b) minimum width $w_{min}$, and (c) asymmetric parameter $\Delta$. for $R_{cr} = 10, 15$ and 20 s$^{-1}$ using TWM.}
        \label{fig:5}
    \end{figure}
    In various applications, it is desirable to maximize the particle current. This could be achieved by taking the appropriate values of geometric parameters of the channel. We study the variation of the particle current $J_x$ with the geometric parameters such as the maximum width $w_{max}$, the minimum width $w_{min}$ and the asymmetric parameter $\Delta$ of the channel. In the following results, the variation of $J_x$ with the parameter of interest is studied keeping all other parameters fixed. The simulations are performed for $L=5~\mu$m, $\Delta = 1$, $w_{min} \simeq 0.46~\mu$m and $w_{max} \simeq 3.5~\mu$m. The dependence of the $J_x$ on the geometric parameters for TWM is shown for $R_{cr} = 10$, 15, 20~s$^{-1}$ in Fig.~\ref{fig:5}. The variation in $J_x$ with $w_{max}$ is shown in Fig.~\ref{fig:5}(a). The channel is relatively flat for low values of $w_{max}$ ($w_{max}\simeq w_{min}$). As $w_{max}$ increase the channel become more corrugated resulting in increase in $J_x$. However, for large $w_{max}$, the bacterium spends most of the time in one compartment resulting in fewer number of trajectories passing from one compartment to another resulting in decrease in $J_x$. In  Fig.~\ref{fig:5}(b), we  show the plot of  $J_x$ vs $w_{min}$ of the channel with fixed values of all other parameters. As $w_{min}$ increase, the boundaries become flatter resulting in reduction in $J_x$. The variation with current with $\Delta$ (for $\A = 1$ $\mu$m, $\B = 0.7$ $\mu$m) in Fig.~\ref{fig:5}(c) indicates that the current increases with the asymmetry in the shape of the channel with $J_x = 0$ for a symmetric periodic channel ($\Delta = 0$).

    \begin{figure}[!htbp]
        \centering
        \includegraphics[scale = 0.08]{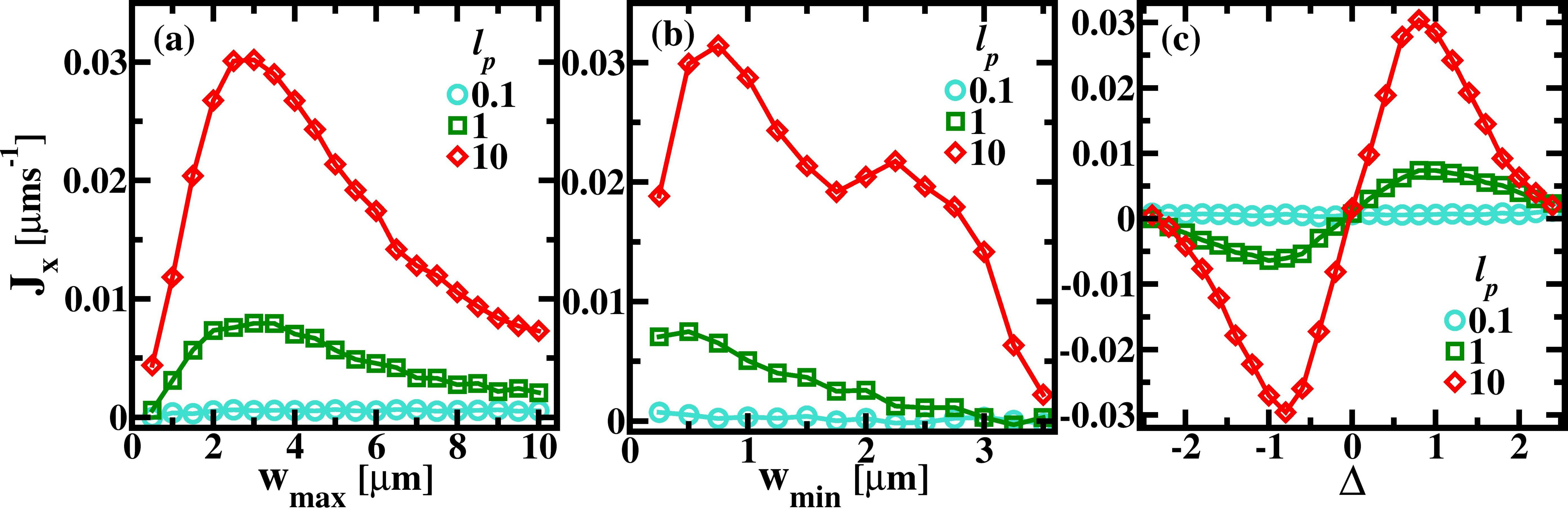}
        \caption{The dependence of the particle current $J_x$ evaluated for the bacterium undergoing RT motility with $\ell_p = 0.1, 1$ and 10 $\mu$m on the channel parameters: (a) maximum width $w_{max}$, (b) minimum width $w_{min}$, and (c) asymmetric parameter $\Delta$.}
        \label{fig:JxVsSPCG}
    \end{figure}
    Next, we report the results for the dependence of $J_x$ on different parameters using the RTM for $\ell_p=0.1$, 1, $10~\mu$m. Fig.~\ref{fig:JxVsSPCG}(a) show the variation of $J_x$ with $w_{max}$. As discussed in the previous paragraph for the case of TWM, there exists a value of $w_{max}$ for which $J_x$ is maximum. In Fig.~\ref{fig:JxVsSPCG}(b), the variation in current as a function of minimum width $w_{min}$ is plotted. The current is less ($J_x\simeq 0$) for small values of $w_{min}$ for a fixed  value of $w_{max}$ ($\simeq 3.5~\mu$m) as only a small number of bacterial trajectories can pass from one compartment to another. The current then starts rising with an increase in $w_{min}$. But as $w_{min}$ increases and approaches the value of $w_{max}$, the channel becomes straighter which reduces the effect of the channel shape on the bacterial dynamics. Hence, the current attains a maximum for a value of $w_{min} = w^*_{min}$ ($w^*_{min}\simeq 0.75~\mu$m for $\ell_p=10~\mu$m) and starts decreasing with further increase in $w_{min}$ beyond this point. The current also seems to have a second maximum for $\ell_p = 10~\mu$m. This merit further examination and analysis in future for understanding it. The current $J_x$ as a function of asymmetric parameter $\Delta$ (with $\A = 1$ $\mu$m, $\B = 0.7$ $\mu$m) is shown in Fig.~\ref{fig:JxVsSPCG}(c). The $J_x\simeq 0$ for $\Delta = 0$ which corresponds to a periodic channel with reflection symmetry. As $\Delta$ increase, the shape of the channel becomes more and more asymmetric, leading to increasing $J_x$. Here, we recall from Eq.~(\ref{eq:delta}) that $\delta(\Delta)$ is an monotonically increasing function for $\Delta > 0$ with $\delta(\Delta)=\delta(-\Delta)$. Therefore, from Eq.~\ref{eq:wmin}, we note that as $\vert\Delta\vert$ increase, $w_{max}$ increases whereas $w_{min}$ decreases resulting in the reduction of current for large $\vert\Delta\vert$.
    
    The dependence of the particle current on the channel parameters are different in the TWM and RTM. This deviation could be due to the difference in the motility near the channel boundaries for the two models. The way SPPs couple with the bounding walls strongly affects its overall dynamics \cite{PM2017}. Also, the difference in the underlying motility mechanisms of different SPPs is known to affect its accumulation of the SPPs at the boundaries \cite{PM2017}. Besides, we could simulate only for a small range of persistence length in TWM due to parameter constraints. The effects of extreme values of persistence length are not observed in TWM as compared to RTM.
    
    \section{Conclusion}\label{sec:4}
    In this work, we study the twitching motility in a 2D asymmetric corrugated channel using the stochastic TWM and the RTM. The TWM quantitatively describes the motility of \ngo resulting from the 2D \textit{tug-of-war} between T4P. On the other hand, RTM is a coarse-grained model constructed by analyzing the experimental trajectories of the bacteria. It is a ubiquitous model to describe the motion of SPPs. The bacterial motion is simulated in the corrugated channel having boundaries with broken reflection symmetry. Due to the confinement of bacteria in the narrow channels, the motility shows anomalous diffusion at different time scales. A non-zero finite value of the mean displacement along the $x$ axis at long time scales signifies that the bacterial motion is rectified and there is a net transport of bacteria in one direction. The probability density and velocity profiles reveal the complex dynamics of the twitching bacteria in the compartment. The plots of spatial probability density show that for a bacterium having persistence length comparable to dimensions of the compartment, it dwells more often near the boundaries than the bulk. The average drift velocity is observed to be higher near the spatial constrictions. The particle current which quantifies this rectification has a dependence on the relative value of the persistence length to the dimensions of the compartment.  The bacteria having the persistence length comparable to the compartment size undergo multiple collision with the boundaries of the channel resulting in a finite particle current along the channel.  We study the variation of particle current for different geometric parameters.  Our simulations reveal that one can optimize the particle current for a given value of persistence length by a suitable selection of the size and the shape of the compartments.
    
    We observe deviations in the features of the motility for the two models. The simulation of the motility using TWM is done for a narrow range of persistence length, which is dictated by the experimental observations. The large values of persistence length taken in the RWM cannot be replicated for TWM due to parameter constraints on TWM. We could simulate the bacterial motility using TWM only for an intermediate value of persistence length. In the overlapping range, the general features of the motility from TWM can be nearly mapped to those from RTM.  However, there are few deviations which we presume could be due to differences in the boundary conditions. In both models, we see enhanced persistence as compared with the motility in the absence of the asymmetric channel. These ratchet effects can also be studied experimentally by creating corrugated channels by micro-printing on a substrate and observing the twitching motility under a microscope \cite{CH2009, GM2009}. We hope that such experiments could further investigate the observations made in our numerical study. We also note that the motility of twitching bacteria in the presence of obstacles may further enhance the persistence, work in this direction is currently underway.
    \acknowledgments
    
    K.B. acknowledges CSIR (IN) for financial support under Grant No. 09/086(1208)/2015-EMR-I. The authors thank IIT Delhi HPC facility for computational resources.
    \bibliographystyle{apsrev4-1}
    \bibliography{ref9719}
\end{document}